\documentclass[12pt]{article}
 \pdfoutput=1
\usepackage{epsfig}
\usepackage{amsmath}
\usepackage{color}
\usepackage{booktabs}
\usepackage{amssymb}

\usepackage{amsmath}
\usepackage{setspace}
\usepackage{array}
\usepackage{times}
\usepackage{bm}
\usepackage{graphicx}
\usepackage{latexsym}
\usepackage{color}

\newtheorem{theorem}{Theorem}[section]
\newtheorem{lemma}{Lemma}[section]

\newtheorem{remark}{Remark}[section]
\newtheorem{definition}{Definition}
\def\qed{\rule{2mm}{2mm}}

\newcommand{\csum}[2]{\displaystyle\sum\limits_{\substack{#1}}^{\substack{#2}}}
\newcommand{\cprod}[2]{\displaystyle\prod\limits_{#1}^{#2}}

\newcommand{\ind}[1]{1_{\{ #1 \}}}
\newcommand{\cond}{\;\middle\vert\;}
\newcommand{\expv}[1] {E\left( #1 \right)}
\newcommand{\pr}[1] {\mathrm{Pr}\left( #1 \right)}

\newcommand{\critA}[1]{\alpha^{(1)}_{#1}}
\newcommand{\critB}[1]{\alpha^{(2)}_{#1}}
\newcommand{\critC}[1]{\alpha^{(3)}_{#1}}
\newcommand{\critD}[1]{\alpha^{(4)}_{#1}}

\newcommand{\maxj}[1] {j_{#1}^*}

\usepackage[dvips, letterpaper, nohead, top=1.3in, bottom=1.3in, left=1.3in, right=1.3in]{geometry}

\begin{document}

\small\normalsize

\title{The Control of the False Discovery Rate in Fixed Sequence Multiple Testing}

\author{
Gavin Lynch \\
Catchpoint Systems, Inc., 228 Park Ave S $\sharp$28080 \\
New York, NY 10003, U.S.A.
\and
Wenge Guo\thanks{The research of
Wenge Guo was supported in part by NSF Grants
DMS-1006021 and DMS-1309162.} \\
Department of Mathematical Sciences \\
New Jersey Institute of Technology \\
Newark, NJ 07102, U.S.A.
\and
Sanat K. Sarkar\thanks{The research of Sanat Sarkar was supported in part
by NSF Grants DMS-1006344 and DMS-1309273.} \\
Department of Statistics, Temple University \\
Philadelphia, PA 19122, U.S.A.
\and
Helmut Finner \\
Institute for Biometrics and Epidemiology \\
German Diabetes Center at the Heinrich-Heine-University \\
Auf'm Hennekamp 65, D-40225 D$\ddot{u}$sseldorf, Germany
}

\date{}
\maketitle

\newpage

\begin{abstract}
Controlling the false discovery rate (FDR) is a powerful approach to
multiple testing. In many applications, the tested hypotheses have
an inherent hierarchical structure. In this paper, we focus on the
fixed sequence structure where the testing order of the hypotheses
has been strictly specified in advance. We are motivated to study
such a structure, since it is the most basic of hierarchical
structures, yet it is often seen in real applications such as statistical
process control and streaming data analysis. We first
consider a conventional fixed sequence method that stops testing
once an acceptance occurs, and develop such a method controlling the
FDR under both arbitrary and negative dependencies. The method under
arbitrary dependency is shown to be unimprovable without losing
control of the FDR and unlike existing FDR methods; it cannot be
improved even by restricting to the usual positive regression
dependence on subset (PRDS) condition. To account for any potential mistakes
in the ordering of the tests, we extend the conventional fixed
sequence method to one that allows  more but a given number of
acceptances. Simulation studies show that the proposed procedures
can be powerful alternatives to existing FDR controlling procedures.
The proposed procedures are illustrated through a real data set from
a microarray experiment.
\end{abstract}


\noindent AMS 2000 subject classifications: Primary 62J15

\noindent KEY WORDS: Arbitrary dependence, false discovery rate, fixed sequence,
multiple testing, negative association, PRDS property, $p$-values.

\section {Introduction}

In many applications of multiple testing, such as genomic research,
clinical trials, and statistical process control, the hypotheses are so
structured that they are to be tested in a particular sequence. This
structure may be a natural one, as in Goeman and Mansmann (2008), where Gene
Ontology imposes a directed acyclic graph structure onto the tested
hypotheses, or can be formed by using a data-driven approach for specifying
the testing order of the hypotheses, as in Kropf and L$\ddot{a}$uter (2002), Kropf et al.
(2004), Westfall et al. (2004), Hommel and Kropf (2005), Finos and Farcomeni (2011), etc. In some applications, it is not even possible
to use the conventional $p$-value based multiple testing methods,
because of some inherent structure among the tested hypotheses. For
example, the hypotheses associated with stream data in sequential
change detection problems (Ross et al., 2011) have a natural temporal
structure, but none of conventional methods, such as the stepwise
procedures, which are applicable only when all of $p$-values are available, can be used here since
the decision concerning a hypothesis has to be made even before the
data associated with the remaining hypotheses are observed.

Some progress has been made in testing structured hypotheses.
However, it has been primarily focused on controlling the familywise
error rate (FWER) (Maurer et al., 1995;
Westfall and Krishen, 2001; Wiens, 2003; Wiens and Dmitrienko, 2005; Hommel et al., 2007;
Huque and Alosh, 2008; Li and Mehrotra, 2008; Rosenbaum, 2008; Wiens and Dmitrienko,
2010; Millen and Dmitrienko, 2011; Dmitrienko et al., 2013). There are few results
towards controlling the false discovery rate (FDR) while accounting
for the structure of the tested hypotheses. Benjamini and Heller (2007), Heller et al. (2009), and
Mehrotra and Heyse (2004) developed methods
for testing hypotheses with a specific hierarchical structure where the
structure is limited to only two levels. Yekutieli (2008)
discussed a method that controls the FDR when the tested hypotheses
have a general hierarchical structure. However, that method is shown
to control the FDR only under independence.

The primary objective of this paper is to help advance the theory and
methods on controlling the FDR for testing structured hypotheses. We
do so by focusing on a structure where the
hypotheses have a fixed pre-defined testing order since this is the simplest of hierarchical
structures, yet it is often seen in real applications such as clinical trials, statistical
process control and streaming data analysis. For such a structure, we will refer to
it as a fixed sequence structure throughout this paper. Very recently,
several methods have been introduced for controlling FDR while testing
pre-ordered hypotheses. Farcomeni and Finos (2013)
developed a `single-step' FDR controlling method for testing hypotheses with the same critical value
$\alpha$, which tests each hypothesis at level $\alpha$ until a stopping condition is reached.
Barber and Candes (2015), G'Sell et al. (2016) and Li and Barber (2016) developed several different
`step-up' FDR controlling procedures in the context of high-dimensional regression for testing hypotheses with fixed sequence structure for which hypotheses are tested from highest-ranked to lowest-ranked, and Lei and Withian (2016) performed asymptotic power analysis for such `step-up' procedures. In addition,
Javanmard and Montanari (2015) developed procedures for controlling the FDR in an online manner
while testing a sequence of possibly infinite pre-ordered hypotheses.

In this paper, we develop `step-down' FDR controlling methods that fully exploit
the fixed sequence structural information, in which hypotheses are tested from lowest-ranked to
highest-ranked. We first consider a
conventional fixed sequence multiple testing method that keeps
rejecting until an acceptance occurs and develop such a method
controlling the FDR under arbitrary dependence. It is shown to be
optimal in the sense that it cannot be improved by increasing even
one of its critical values without losing control over the FDR, or
even by imposing a positive dependence condition on the $p$-values,
such as the standard PRDS (positive regression dependence on subset)
condition of Benjamini and
Yekutieli (2001). This is different from what
happens in case of non-fixed sequence multiple testing. For
instance, the so-called BY method of Benjamini and
Yekutieli (2001) that
controls the FDR under arbitrary dependence can be improved
significantly by the BH method of Benjamini and Hochberg (1995) by imposing
this PRDS condition. Since our procedure cannot be improved under
positive dependence, we consider the case of negative dependence and
develop a more powerful conventional fixed sequence multiple testing
method controlling the FDR under negative dependence which includes
independence as a special case.

There is a potential for loss of power in a conventional fixed
sequence multiple testing method if the ordering of the hypotheses,
particularly for the earlier ones, does not match with that of their
true effect sizes, potentially leading to some earlier hypothesis
being accepted and the follow-up hypotheses having no chance to be
tested. To mitigate that, we consider generalizing the conventional
fixed sequence multiple testing to one that allows more than one but
a pre-specified number of acceptances, and develop such generalized
fixed sequence multiple testing methods controlling the FDR under
both arbitrary dependence and independence.

It is not always the case in real data applications that the
hypotheses will have a natural fixed sequence structure or
information about how to order them will be available a priori.
Nevertheless, the data itself can often provide information on how
to order the hypotheses. In this paper, we discuss such a
data-driven ordering strategy which can be applied to a broad
spectrum of multiple testing problems such as one-sample and
two-sample t-tests, and one-sample and two-sample nonparametric
tests. Through simulation studies and a real microarray data
analysis, this strategy coupled with our proposed fixed sequence
methods is seen to perform favorably against the corresponding
non-fixed sequence methods under certain settings.

The paper is organized as follows. With some concepts and background
information given in Section 2, we present the developments of our
conventional and generalized fixed sequence procedures controlling
the FDR under various dependencies in Sections 3 and 4,
respectively. Our fixed sequence procedures coupled with a data-driven ordering strategy for the hypotheses are applied to a real microarray data in Section 5. The findings from some
simulation studies on the performances of our procedures are given
in Section 6. Some concluding remarks are made in Section 7 and
proofs of some results are given in the Appendix.

\section{Preliminaries}

Suppose that $H_i, i = 1, \ldots, m$, are the $m$ null hypotheses
that are ordered a priori and are to be simultaneously tested based
on their respective $p$-values $P_i, i = 1, \ldots, m$. Let $m_0$
and $m_1$ of these null hypotheses be true and false, respectively.
For notational convenience, we denote the index of the $i$th true
null hypothesis by $u_i$ and the set of indices of the true null
hypotheses by $I_0$. Let $V$ and $S$ be the numbers of true and
false null hypotheses, respectively, among the $R$ rejected null
hypotheses in a multiple testing procedure. Then, the familywise
error rate (FWER) and false discovery rate (FDR) of this procedure
are defined respectively as $\text{FWER} = \text{Pr}(V > 0)$ and
$\text{FDR} = E\left (\frac{V}{\max(R,1)} \right ) = E\left(
\frac{V}{V+S}\ind{V>0} \right) $.

Typically, the hypotheses are ordered based on their $p$-values and
multiple testing is carried out using a stepwise or single-step
procedure. However, when these hypotheses are ordered a prior and not according to their $p$-values,
multiple testing is often performed using a fixed sequence method. Given a non-decreasing sequence of
critical constants $0 < \alpha_1 \le \ldots \le \alpha_m$, a
conventional fixed sequence method is defined as follows:
\begin{definition} \emph{(Conventional fixed sequence method)}
\begin{enumerate}
    \item If $P_1 \le \alpha_1$, then reject $H_1$
    and continue to test $H_2$; otherwise, stop.
    \item If $P_i \le \alpha_i$ then reject $H_i$
    and continue to test $H_{i+1}$; otherwise, stop.
\end{enumerate}
\end{definition}

Thus, a conventional fixed sequence method continues testing in the
pre-determined order as long as rejections occur. Once an acceptance
occurs, it stops testing the remaining hypotheses. In Section 4, we
will generalize a conventional fixed sequence method to allow a
given number of acceptances. It should be noted that a conventional
fixed sequence method with common critical constant $\alpha$, which
is often called the fixed sequence procedure in the literature,
strongly controls the FWER at level $\alpha$ (Maurer et al., 1995). We
will refer to it as the FWER fixed sequence procedure in this paper
in order to distinguish it from other fixed sequence methods
designed to control the FDR.

Regarding assumptions we make about the $p$-values in this paper, we assume that the true null $p$-values, which we denote for notational convenience by $\widehat P_i$, for $i =1, \ldots, m_0$,
are marginally distributed as follows:
\begin{equation}
\pr{\widehat P_i \le p} \le p \text{~~~for~any~} p \in (0, 1). \label{EQN_UNIFORM}
\end{equation}
One of several types of dependence, like arbitrary dependence,
positive dependence, negative dependence, and independence, has been
assumed to characterize a dependence structure among the $p$-values.

By arbitrary dependence, we mean that the $p$-values do not have any
specific form of dependence. The positive dependence is
characterized by the positive regression dependence on subset (PRDS)
property (Benjamini and Yekutieli, 2001) as defined below:

\begin{definition} (PRDS) The vector of $p$-values  $\vec{P}$ is PRDS on
the vector of null $p$-values $\vec{P_0} = (\widehat P_1, \ldots,
\widehat P_{m_0})$ if for every increasing set $D$ and for each $i
=1, \ldots, m_0$, the  conditional probability $\pr{\vec{P} \in D
\cond \widehat P_i = p}$ is non-decreasing in $p$. \end {definition}

Several multivariate distributions possess this property (see, for
instance, Benjamini and
Yekutieli, 2001; Sarkar, 2002). The negative
dependence is characterized by the following property:
\begin {definition} \label{NEGATIVE_ASSOCIATION} (Negative
Association) The vector of $p$-values  $\vec{P}$ is negatively
associated with null $p$-values if for each $i =1, \ldots, m_0$, the
following inequality holds:
\begin{eqnarray}
& & ~~ \pr{\widehat P_i \le p_{u_i}, P_j \le p_j, j = 1, \ldots, m,
\text{~with~} j \neq u_i} \nonumber \\
& \le & \pr{\widehat P_i \le p_{u_i}} \pr{P_j \le p_j, j = 1,
\ldots, m, \text{~with~} j \neq u_i}, \label{EQN_NEG_ASSOC}
\end{eqnarray}
for all fixed $p_j$'s.
\end{definition}

Several multivariate distributions posses the conventional negative
association property, including multivariate normal with
non-positive correlation, multinomial, dirichlet, and multivariate
hypergeometric (Joag Dev and Proschan, 1983). It is easily seen that independence is a special case of negative dependence.

\section{Conventional Fixed Sequence Procedures} \label{SUBSECTION_FIX_SEQ_NO_ACCEPT}

In this section, we present the developments of two simple
conventional fixed sequence procedures controlling the FDR under
both arbitrary dependence and negative dependence conditions on the
$p$-values.

\subsection{Procedure under arbitrary dependence}

Since the FDR is more liberal than the FWER, a conventional fixed sequence
method controlling the FDR under arbitrary dependence is expected to
have critical values that are at least as large as $\alpha$, the common critical constant
for the FWER fixed sequence method. In the following, we present
such a simple conventional fixed sequence FDR controlling procedure.
\begin{theorem} \label{THM_ANY_JOINT_FIX_SEQ_NO_ACC}
Consider a conventional fixed sequence procedure with critical
constants
\[
\critA{i} = \min\left(\frac{m\alpha}{m-i+1}, 1\right), \; i = 1, \dots,
m. \nonumber
\]
\begin{itemize}
    \item[(i)]This procedure
strongly controls the FDR at level $\alpha$ under arbitrary
dependence of the $p$-values.
    \item[(ii)] One cannot increase even one of the critical constants
    $\critA{i}, i = 1, \dots, m,$ while keeping the remaining fixed without losing control of the FDR.
    This is true even when $\vec{P}$ is assumed to be PRDS on $\vec{P}_0$.
\end{itemize}
\end{theorem}

Proof of (i). Since $u_1$ is the index of the first true null
hypothesis, the first $u_1-1$ null hypotheses are all false.
Note that the event $\{V>0\}$ implies that $S \ge u_1-1$ and $\hat{P}_1
\le \critA{u_1}$, and therefore we have
\begin{align*}
\text{FDR} &= E\left( \frac{V}{V+S}\ind{V>0} \right) \le E\left( \frac{m_0}{m_0+u_1-1}\ind{V>0} \right) \\
&= \frac{m_0}{m_0+u_1-1}Pr(V>0) \le \frac{m_0}{m_0+u_1-1}Pr(\hat{P}_1 \le \critA{u_1}) \\
&\le \frac{m-u_1+1}{m}\critA{u_1} \le \alpha.
\end{align*}
The first inequality follows from the fact that $V/(V+S)$ is an
increasing function of $V$ and a decreasing function of $S$.  The
third inequality follows from the fact that $m_0/(m_0+u_1-1)$ is an
increasing function of $m_0$ and $m_0 \le m-u_1+1$ since there are
at least $u_1-1$ false nulls. This proves part (i).

For a proof of part (ii), see Appendix. \qed

\begin{remark} \rm
Theorem \ref{THM_ANY_JOINT_FIX_SEQ_NO_ACC} shows that when
controlling the FDR under arbitrary dependence, the operating
characteristic of the proposed fixed sequence method is much
different from that of the usual stepwise procedure of Benjamini and
Yekutieli (Benjamini and Yekutieli, 2001) that relys on $p$-value based
ordering of the hypotheses. It cannot be further improved, even by
imposing the PRDS assumption on the $p$-values, unlike the BY
procedure that is known to be greatly improved by the
Benjamini-Hochberg (BH) procedure under such positive dependence.
Also, as shown in our proof of Theorem
\ref{THM_ANY_JOINT_FIX_SEQ_NO_ACC}(ii) under arbitrary dependence
(see Appendix), the least favorable configuration (the configuration
which leads to the largest error rate, see Finner and Roters, 2001)
of the newly suggested fixed sequence FDR controlling procedure is
when the ordering of the hypotheses is perfect (i.e, when all the
false null hypotheses are tested before the true ones), the false
null $p$-values are all $0$ with probability 1, and the true null
$p$-values are the same with each following $U(0,1)$ distribution.
This least favorable configuration is much different from that given
in Guo and Rao (2008) for the BY procedure under arbitrary dependence.

Although the procedure in Theorem \ref{THM_ANY_JOINT_FIX_SEQ_NO_ACC}
cannot be improved under the PRDS condition, we consider in the next
subsection the condition of negative dependence which includes
independence as a special case, and under such condition, develop a
more powerful conventional fixed sequence method that controls the
FDR.
\end{remark}

\subsection{Procedure under negative dependence}

When the $p$-values are negatively associated as defined in Section
2, the critical constants of the conventional fixed sequence
procedure in Theorem \ref{THM_ANY_JOINT_FIX_SEQ_NO_ACC} can be
further improved as in the following:

\begin{theorem} \label{THM_INDEPENDENCE_FIX_SEQ_NO_ACC}
The conventional fixed
sequence method with critical constants
\[
\critB{i} = \frac{i \alpha}{1+(i-1)\alpha}, i = 1, \ldots, m
\]
strongly controls the FDR at level $\alpha$ when the $p$-values are
negatively associated on the true null $p$-values.
\end{theorem}

To prove Theorem \ref{THM_INDEPENDENCE_FIX_SEQ_NO_ACC}, we use the
following lemma, with proof given in Appendix:
\begin{lemma} \label{LEMMA_FDR_PROCESS_SPECIAL}
Let $m_{0,i}$ and $m_{1,i}$ respectively denote the numbers of true
and false null hypotheses among the first $i$ null hypotheses, and
\begin{align*}
d_i =
    \begin{cases}
    \ind{i \in I_0} &\text{ if }~ i = 1 \\
    (m_{1,i-1}\ind{i \in I_0}-m_{0,i-1}\ind{i \notin I_0})/(i(i-1)) &\text{ if }~ i > 1. \\
    \end{cases}
\end{align*}
Then, the FDR of any fixed sequence procedure can be expressed as
\[
\emph{FDR} = \csum{i = 1}{m}d_i\pr{R \ge i}.
\]
\end{lemma}

Proof of Theorem \ref{THM_INDEPENDENCE_FIX_SEQ_NO_ACC}. If
$u_1 = 1$, then
\[
\text{FDR} \le \text{FWER} = \pr{V > 0} = \pr{\widehat P_1 \le
\critB{1}} \le \alpha.
\]

If $u_1 > 1$, then by Lemma \ref{LEMMA_FDR_PROCESS_SPECIAL},
\begin{align}
\text{FDR} &= \csum{i=1}{m}d_i \pr{R \ge i} = \csum{i=u_1}{m}d_i \pr{R \ge i} \nonumber \\
&= \csum{i=u_1}{m}\left(d_i + \frac{m_{1,i}\alpha}{i}\right) \pr{R
\ge i} - \csum{i=u_1}{m}\frac{m_{1,i}\alpha}{i} \pr{R \ge i}.
\label{THM_INDEPENDENCE_FIX_SEQ_NO_ACC_EQN1}
\end{align}
The second equality follows from the fact that $d_i = 0$ for $i = 1,
\dots, u_1-1$.

For each $i = u_1, \dots, m$, the following inequality holds.
\begin{equation}
\left(d_i + \frac{m_{1,i}\alpha}{i}\right) \pr{R \ge i} \le
\frac{m_{1,i-1}\alpha}{i-1}Pr(R \ge i-1).
\label{THM_INDEPENDENCE_FIX_SEQ_NO_ACC_EQN2}
\end{equation}
To see this, we consider, separately, the case when $i \in I_0$ and
when $i \notin I_0$.  Suppose $i \in I_0$, then $m_{1,i-1} =
m_{1,i}$ and
\begin {eqnarray*}
&& \left(d_i + \frac{m_{1,i}\alpha}{i}\right) \pr{R \ge i} \\
& = & \left(\frac{m_{1,i-1}}{i(i-1)} +
\frac{m_{1,i-1}\alpha}{i}\right)
\pr{P_1 \le \critB{1}, \ldots, P_{i-1} \le \critB{i-1}, P_i \le \critB{i}} \\
 & \le & \frac{ m_{1,i-1} \left (1 +
(i-1)\alpha \right )}{i(i-1)}
 \pr{P_1 \le \critB{1}, \ldots, P_{i-1} \le \critB{i-1}} \pr{P_i \le \critB{i}} \\
& \le & \frac{m_{1,i-1}\alpha}{i-1}Pr(R \ge i-1).
\end {eqnarray*}
The first and second inequalities follow from (\ref{EQN_NEG_ASSOC}) and (\ref{EQN_UNIFORM}),
respectively.

 Now suppose $i \notin I_0$, then $m_{1,i} = m_{1,i-1} + 1$ and
\begin {eqnarray*}
&& \left(d_i + \frac{m_{1,i}\alpha}{i}\right) \pr{R \ge i} = \left(-\frac{m_{0,i-1}}{i(i-1)} + \frac{(m_{1,i-1}+1)\alpha}{i}\right) \pr{R \ge i} \\
& \le &\left(-\frac{m_{0,i-1}\alpha}{i(i-1)} + \frac{(m_{1,i-1}+1)\alpha}{i}\right) \pr{R \ge i} \\
& = &\frac{m_{1,i-1}\alpha}{i-1} \pr{R \ge i} \le
\frac{m_{1,i-1}\alpha}{i-1} \pr{R \ge i-1}.
\end {eqnarray*}
In the second equality, we use the fact that
$m_{0,i-1}+m_{1,i-1}=i-1$.

Applying (\ref{THM_INDEPENDENCE_FIX_SEQ_NO_ACC_EQN2}) to
(\ref{THM_INDEPENDENCE_FIX_SEQ_NO_ACC_EQN1}), we have
\begin{eqnarray*}
\text{FDR} & = &\csum{i=u_1}{m}\left(d_i + \frac{m_{1,i}\alpha}{i}\right) \pr{R \ge i} - \csum{i=u_1}{m}\frac{m_{1,i}\alpha}{i} \pr{R \ge i} \nonumber \\
 & \le & \csum{i=u_1}{m}\frac{m_{1,i-1}\alpha}{i-1} \pr{R \ge i-1} - \csum{i=u_1}{m}\frac{m_{1,i}\alpha}{i} \pr{R \ge i} \\
& = & \alpha \pr{R \ge u_1-1} - \frac{m_1\alpha}{m}\pr{R = m} \\
& \le \alpha.
\end{eqnarray*}
The equality follows from that fact that $m_{1,u_1-1} = u_1-1$,
since the first $u_1-1$ hypotheses are false. \qed

\begin{remark} \rm
The conventional fixed sequence procedure in Theorem
\ref{THM_INDEPENDENCE_FIX_SEQ_NO_ACC} is nearly optimal in the sense
that the upper bound of the FDR of this procedure is very close to
$\alpha$ under certain configurations. Consider the following
configuration: All the false null hypotheses are tested before the
true null hypotheses, the false null $p$-values are all $0$ with
probability 1, and the true null $p$-values are independently
distributed as $U(0,1)$. Under this configuration, it is easy to check that
(\ref{THM_INDEPENDENCE_FIX_SEQ_NO_ACC_EQN2}) becomes an equality.
Following the proof of Theorem
\ref{THM_INDEPENDENCE_FIX_SEQ_NO_ACC}, the FDR of this procedure is
exactly $\alpha - \frac{m_1\alpha}{m}\pr{R = m}$. When $m_1/m$
approaches to $\pi_1$ as $m \rightarrow \infty$ with $0 \le \pi_1 <
1$, an approximate lower bound of the FDR is $\alpha - \pi_1 \alpha
e^{-(1-\pi_1)\frac{1-\alpha}{\alpha}}$.

To see why, we first note that
\[
\frac{m_1\alpha}{m}\pr{R = m} =
\frac{m_1\alpha}{m}\cprod{i=m_1+1}{m}\critB{i} <
\frac{m_1\alpha}{m}(\critB{m})^{m-m_1}.
\]
Then, by simple calculation, we have
\begin {eqnarray*}
&& \lim_{m \rightarrow \infty} \text{FDR} \ge \alpha - \lim_{m
\rightarrow
\infty} \frac{m_1\alpha}{m}(\critB{m})^{m-m_1} \\
& = & \alpha - \lim_{m \rightarrow \infty}
\frac{m_1\alpha}{m}\left(1 +
\frac{1-\alpha}{\alpha}\frac{1}{m}\right)^{-m(1-\frac{m_1}{m})} \\
& = & \alpha - \pi_1\alpha e^{-(1-\pi_1)\frac{1-\alpha}{\alpha}}.
\end {eqnarray*}
This lower bound of the FDR is very close to the pre-specified level
$\alpha$. For example, for large $m$, if $m_1/m = 0.2$, then with
$\alpha = 0.05$, the lower bound under the above configuration is
about 0.04999975.
\end{remark}

\begin{remark} \rm
Figure \ref{IMG_CRITICAL_VALUE_PLOT} compares the critical constants
for the proposed procedures along with the FWER fixed sequence
procedure at level $\alpha$. It should be noted that the first few
critical constants are the most important ones. If the first few
values are too small, then the procedure might stop too early and
the remaining hypotheses will not have a chance to be tested. With
this in mind, it can be seen that the critical constants of the
procedure introduced in Theorem
\ref{THM_INDEPENDENCE_FIX_SEQ_NO_ACC} are by far the best, and the
critical constants of the procedure in Theorem
\ref{THM_ANY_JOINT_FIX_SEQ_NO_ACC} are slightly better than those of
the conventional fixed sequence procedure.
\begin{figure}
\begin{center}
  \includegraphics[width=0.375\textwidth,angle=-90]{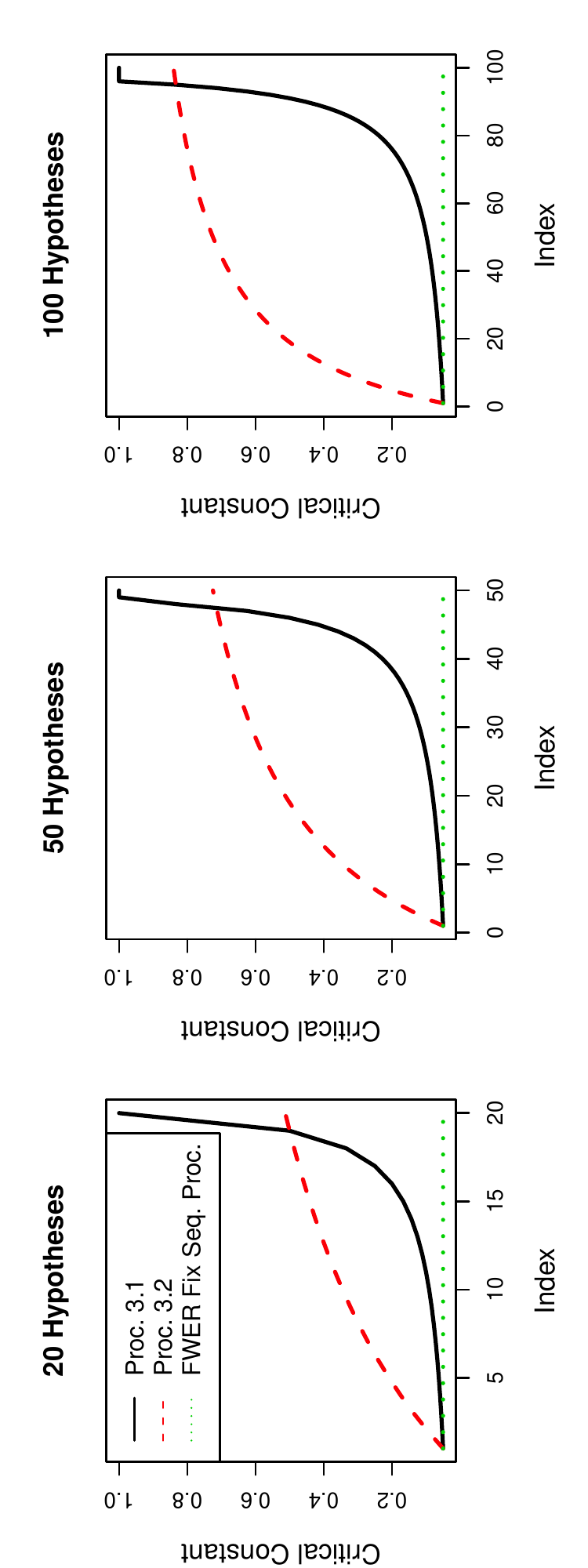}
  \vskip -10pt
  \caption{\textit{A plot of the critical constants of the procedures in Theorems
  \ref{THM_ANY_JOINT_FIX_SEQ_NO_ACC} (solid line), \ref{THM_INDEPENDENCE_FIX_SEQ_NO_ACC}
  (dashed line), and the FWER fixed sequence procedure (dotted line)
  for $m=$20, 50, and 100.}}
  \label{IMG_CRITICAL_VALUE_PLOT}
\end{center}
\end{figure}
\end{remark}

\section{Fixed Sequence Procedures that Allow More Acceptances}

A conventional fixed sequence method might potentially lose power if
an early null hypothesis fails to be rejected, with the remaining
hypotheses having no chance of being tested. To remedy this, we
generalize a conventional fixed sequence method to one that allows a
certain number of acceptances.  The procedure will keep testing
hypotheses until a pre-specified number of acceptance has been
reached. The same idea has also been used by Hommel and
Kropf (2005) to develop FWER controlling procedures in fixed-sequence multiple testing.

Suppose $k$ is a pre-specified positive integer and $\alpha_1 \le \cdots \le \alpha_m$
is a non-decreasing sequence of critical constants. A fixed sequence method
that allows more acceptances is defined below.
\begin{definition} \emph{(Fixed sequence method stopping on the $k^{th}$ acceptance)}
\begin{enumerate}
    \item If $P_1 \le \alpha_1$, then reject $H_1$; otherwise, accept $H_1$.
    If $k > 1$ or $H_1$ is rejected, then continue to test $H_2$; otherwise stop.
    \item If $P_i \le \alpha_i$, then reject $H_i$; otherwise, accept $H_i$.
    If the number of accepted hypotheses so far is less than $k$, then continue
    to test $H_{i+1}$; otherwise stop.
\end{enumerate}
\end{definition}

It is easy to see that when $k = 1$, the fixed
sequence method stopping on the $k^{th}$ acceptance reduces to the conventional one.

\begin{theorem} \label{THM_ANY_JOINT_FIX_SEQ_WITH_ACC}
The fixed sequence method stopping on the $k^{th}$ acceptance with
critical constants
\begin{align*}
\critC{i} =
\begin{cases}
\frac{\alpha}{k} &\text{ if } \quad i = 1, \dots, k \\
\frac{(m - k + 1)\alpha}{(m-i+1)k} &\text{ if } \quad i = k+1, \dots, m
\end{cases}
\end{align*}
strongly controls the FDR at level $\alpha$ under arbitrary
dependence of the $p$-values.
\end{theorem}

Proof. Let $U$ be the index of the first rejected true null hypothesis.  If
no true null hypotheses are rejected, then set $U = 0$.  We will show
that for $i = 1, \ldots, m_0$,
\begin{align}
\expv{\frac{V}{V+S}\ind{U = u_i}} \le \frac{\alpha}{k}.
\label{THM_ANY_JOINT_FIX_SEQ_WITH_ACC_EQN1}
\end{align}

If $i \le k$, then
\[
\expv{\frac{V}{V+S}\ind{U = u_i}} \le \pr{U=u_i} \le \pr{\hat{P}_i
\le \frac{\alpha}{k}} \le \frac{\alpha}{k}.
\]

Now, assume $i > k$. Note that the event $\{U = u_i \}$ implies $V
\le m-u_i+1$ and $S \ge u_i-k$, because the first $u_i-1$ hypotheses
were either false nulls or not rejected true nulls and among the
first $u_i-1$ hypotheses tested, there can be at most $k-1$
acceptances. Thus,
\begin{eqnarray*}
    & &\expv{\frac{V}{V+S}\ind{U = u_i}}
\le  \expv{\frac{m - u_i + 1}{(m-u_i+1)+(u_i-k)}\ind{U = u_i}} \\
 & \le & \frac{m - u_i + 1}{m-k+1}\pr{\hat{P}_i \le \critC{u_i}} \le
\frac{\alpha}{k}.
\end{eqnarray*}
The first inequality follows from the fact that $V/(V+S)$ is an
increasing function of $V$ and a decreasing function of $S$.

From (\ref{THM_ANY_JOINT_FIX_SEQ_WITH_ACC_EQN1}), we have
\[
\text{FDR} = \csum{i = 1}{\min(m_0, k)}\expv{\frac{V}{S+V}\ind{U =
u_i}} \le \csum{i = 1}{k}\frac{\alpha}{k} = \alpha,
\]
where the first equality follows from the fact that if
none of the first $k$ true null hypotheses are rejected,
then $V = 0$. ~\qed

\vskip 5pt

We should point out that the result in Theorem \ref{THM_ANY_JOINT_FIX_SEQ_WITH_ACC} is weaker than that in
Theorem \ref{THM_ANY_JOINT_FIX_SEQ_NO_ACC}, although the method in Theorem \ref{THM_ANY_JOINT_FIX_SEQ_WITH_ACC} reduces to that in Theorem \ref{THM_ANY_JOINT_FIX_SEQ_NO_ACC} when $k=1$. More specifically, we cannot prove that the
procedure in Theorem \ref{THM_ANY_JOINT_FIX_SEQ_WITH_ACC} is optimal in the sense that its critical
constants cannot be further improved without losing control of the
FDR under arbitrary dependence of the $p$-values. However, under certain distributional assumptions on the $p$-values,
the critical constants of this procedure can potentially
be improved. In particular, we have the following result.

\begin{theorem} \label{THM_INDEPENDENCE_FIX_SEQ_WITH_ACC}
Consider a fixed sequence method stopping on the $k^{th}$ acceptance
with critical constants
\[
\critD{i} = \frac{(r_{i-1}+1)\alpha}{k + (i-k)\alpha}, i = 1, \dots,
m,
\]
where $r_i$ is the number of rejections among the first $i$ tested
hypotheses, with $r_0 = 0$, for $i = 1, \ldots, m$. This
procedure strongly controls the FDR at level $\alpha$ if the true null $p$-values are mutually independent and are independent of the false null $p$-values.
\end{theorem}

Before presenting a proof of the above theorem, let us introduce
a few more notations.
For $i = 1, \ldots, m$, let
$V_i$ and $S_i$ be the numbers of false rejections and true
rejections among the first $i$ rejections and $J_i$ be the index of
the $i^{th}$ rejected hypothesis. If there are less than $i$
rejections, we define $V_i = V_{i-1}$, $S_i = S_{i-1}$ and $J_i =
m+1$. In addition, for notational convenience, define $V_0 = S_0 =
J_0 = 0, V_0/0 = 0,$ and $S_0/0 = 1$.

We use the following two lemmas, with proofs
given in Appendix, to prove the theorem.
\begin{lemma} \label{LEMMA_FDR_PROCESS}
The FDR of any fixed sequence method stopping on the $k^{th}$
acceptance can be expressed as
\[
\emph{FDR} = \expv{\csum{i =
1}{m}\left(\frac{V_i}{i}-\frac{V_{i-1}}{i-1}\right)\ind{J_i < m+1}}.
\]
\end{lemma}

\vskip 5pt

\begin{lemma} \label{LEMMA_INDEP2}
Consider the procedure defined in Theorem
\ref{THM_INDEPENDENCE_FIX_SEQ_WITH_ACC}, the following inequality
holds for $i = 1, \dots, m$,
\begin{eqnarray}
& & \expv{\left(\frac{V_i}{i}-\frac{V_{i-1}}{i-1}\right)\ind{J_i < m+1}} \nonumber \\
& \le &
\expv{\frac{(k-J_{i-1}+i-1)\alpha}{k}\frac{S_{i-1}}{i-1}\ind{J_{i-1}
< m+1} - \frac{(k-J_i+i)\alpha}{k}\frac{S_i}{i}\ind{J_i < m+1}}.
\nonumber
\end{eqnarray}
\end{lemma}

\vskip 8pt

Proof of Theorem \ref{THM_INDEPENDENCE_FIX_SEQ_WITH_ACC}. By
Lemmas \ref{LEMMA_FDR_PROCESS} and \ref{LEMMA_INDEP2}, we have
\begin{eqnarray*}
    \text{FDR} & = & \expv{\csum{i = 1}{m}\left(\frac{V_i}{i}-\frac{V_{i-1}}{i-1}\right)\ind{J_i < m+1}} \\
    & \le &  \expv{\alpha - \frac{(k-J_1+1)\alpha}{k}S_1\ind{J_i < m+1}} \\
    & & + ~E\left(\csum{i = 2}{m}\left(\frac{(k-J_{i-1}+i-1)\alpha}{k}\frac{S_{i-1}}{i-1}\ind{J_{i-1} < m+1} \right. \right. \\
     & & \left. \left. - ~\frac{(k-J_i+i)\alpha}{k}\frac{S_i}{i}\ind{J_i < m+1}\right) \right)\\
 & = & \expv{\alpha - \frac{(k-J_m+m)\alpha}{k}\frac{S_m}{m}\ind{J_m <
m+1}} \le \alpha. ~\qed
\end{eqnarray*}

\begin{remark} \rm
When $k = 1$, the generalized fixed sequence method in the above
theorem reduces to the conventional fixed sequence method in
Theorem \ref{THM_INDEPENDENCE_FIX_SEQ_NO_ACC}, since in this case $r_{i-1} = i-1$, and thus continues to control the FDR when the $p$-values are negatively associated. However, when $k > 1$, it can only control the FDR under the independence assumption made in the theorem. It can be shown that this method, when $k >1$, may no longer control the FDR when the $p$-values are negatively associated. Consider, for example, the problem of
simultaneously testing two hypotheses $H_1$ and $H_2$ for which both of them
are true, the associated $p$-values $\widehat P_1$ and $\widehat
P_2$ are both $U(0, 1)$, and $\widehat P_2 = 1 - \widehat P_1$. It
is easy to see that under such configuration, when $k = 2$, the FDR of this
procedure is equal to $\frac{\alpha}{2 - \alpha} + \frac{\alpha}{2}
> \alpha.$
\end{remark}

\section{Data Driven Ordering}

The applicability of the aforementioned fixed sequence methods depends
on availability of natural ordering structure among the hypotheses.
When the hypotheses cannot be pre-ordered,
one can use pilot data available
to establish a good ordering among the hypotheses in some cases. For example, in
replicated studies, the hypotheses for the follow-up study can be
ordered using the data from the primary study. However, when prior information is unavailable ordering information can usually be assessed from the data itself. Such data-driven ordering has been used by several authors in fixed sequence methods controlling the FWER and generalized FWER (Kropf
and L$\ddot{a}$uter, 2002; Kropf et al., 2004; Westfall et al., 2004; Hommel and Kropf, 2005; Finos
and Farcomeni, 2011).

Assume that the variables of interest are $X_{i}$, $i = 1, \ldots, m$, with $n$ independent observations $X_{i1}, \ldots, X_{in}$ on each $X_i$. An ordering statistic, $Y_i$, and a test statistic, $T_i$,  are determined for each $i = 1, \dots, m$. The $Y_i$'s are used to order all of the hypotheses $H_1, \ldots, H_m$, $T_i$ is used to test the hypothesis $H_i, i=1, \ldots, m$, and $P_i$ is the corresponding $p$-value. In addition, $Y_i$ is chosen such that it is independent of the $T_i$'s under $H_i$ and tends to be larger as the effect size increases. The approach is outlined below.
\begin{definition} Data-Driven Ordering Procedure
\begin{enumerate}
\item The hypotheses are ordered based on $Y_1, \dots, Y_m$ where the hypothesis corresponding to the largest $Y_i$ is placed first, the hypothesis corresponding to the second largest is placed second, and so on.
\item The hypotheses are tested using a fixed sequence procedure based on the the $p$-values $P_1, \ldots, P_m$ and the testing order established in Step 1.
\end{enumerate}
\end{definition}

We give a few examples to further illustrate the approach.

\vspace{10pt}
\emph{Example 1: One sample T-test.} Consider testing $H_i: \mu_i = 0$ against $H_i': \mu_i \neq 0$ simultaneously where $X_i$ follows a $N(\mu_i, \sigma^2)$ distribution. Let $\bar{X}_i = \sum_{j=1}^n X_{ij}/n$ and $s_i^2 = \sum_{j=1}^{n}(X_{ij}-\bar{X}_i)^2/(n-1)$ be
the sample mean and variance, respectively,  based on the
observations $X_{i1}, \ldots, X_{in}$. Let $Y_i = \sum_{j=1}^n X_{ij}^2$ be the ordering statistics, that is, the hypotheses are ordered according to the values of the corresponding sums of squares,
and $T_i = \sqrt{n}\bar{X}_i/s_i$ is the usual $t$-test statistic for testing $H_i$. Then, $P_i = 2 \left(1 -F(|T_i|) \right)$, $i=1, \ldots, m$, where $F(\cdot)$ is the cdf of the $t$-distribution with $n-1$ degrees
of freedom, are the $p$-values.  When $\mu_i = 0$, $T_i$ and $Y_i$ are independent (see, for instance, Lehmann and Romano, 2005; p. 156). Furthermore,
\[
E (Y_i) = n(\mu_i^2 + \sigma^2),
\]
which suggests that $|\mu_i|$ tends to increase as $Y_i$ increases.

\vspace{10pt}
\emph{Example 2: Two sample T-test.} Consider testing $H_i: \mu^{(1)}_i = \mu^{(2)}_i$ against $H_i': \mu^{(1)}_i \neq \mu^{(2)}_i$ simultaneously using $n = n_1 + n_2$ data vectors. Suppose $X_{ij}^{(l)}$, $j=1, \ldots, n_l$, follows a $N(\mu^{(l)}_i, \sigma^2)$ distribution, for $l = 1, 2$. Then, the hypotheses can be tested using the two-sample $t$-test statistics $T_i$ and are ordered through the values of the `total sum of squares,' which is $Y_i = \sum_{l=1}^2\sum_{j=1}^{n_l} (X_{ij}^{(l)} - \bar X_i)^2$, where $\bar X_i = \sum_{l=1}^2\sum_{j=1}^{n_l} X_{ij}^{(l)}/n$, for $i=1, \ldots, m$. The rationale behind this is independence between the $Y_i$'s and $T_i$ under $H_i$ (see, for instance, Westfall et al., 2004), and the following result: $E [Y_i] = (n-1) \sigma^2 + n_1n_2 (\mu_{i}^{(1)} - \mu_{i}^{(2)})^2 /n$.

\vspace{10pt}
\emph{Example 3: Nonparametric test.} Kropf et al. (2004) describe a data-driven ordering strategy for nonparametric tests. In the one sample case, we are interested in testing $H_i: \mu_i = 0$ against $H_i': \mu_i \neq 0$ where $X_{ij}, j= 1, \ldots, n$ are assumed to be symmetric about $\mu_i$. The hypotheses are tested using the one-sample Wilcoxon test and ordered based on $Y_i = \text{med}(|X_{i1}|, \dots, |X_{in}|)$. In the two sample case, we are interested in testing $H_i: \mu^{(1)}_i = \mu^{(2)}_i$ against $H_i': \mu^{(1)}_i \neq \mu^{(2)}_i$ using $n = n_1 + n_2$ data vectors, where $X^{(l)}_{ij}, j = 1, \ldots, n_l$ are assumed to be symmetric about $\mu^{(l)}_i$ for $l = 1, 2$. The hypotheses are tested using the two-sample Wilcoxon test and ordered based on the interquartile range $Y_i = q_{3i} - q_{1i}$, where $q_{1i}$ and $q_{3i}$ are respectively the $1^{st}$ and $3^{rd}$ quartile of the mixture of $X^{(1)}_{ij}$'s and $X^{(2)}_{ij}$'s.

\vspace{10pt}
When our proposed fixed sequence procedures are used in applications coupled with the aforementioned data-driven ordering strategy, the FDR controls are still maintained under the independence assumption, if the ordering statistics are chosen to be independent of the test statistics in the data-driven ordering strategy, even though the same data is repeatedly used for ordering and testing the hypotheses. We have the following result.

\vspace{10pt}
\begin{theorem} \label{THM_DATA_DRIVEN}
Suppose $X_1, \dots, X_m$ are mutually independent. If the hypotheses $H_1, \dots, H_m$ are ordered based on the ordering statistics $Y_i, i= 1, \ldots, m$, tested using the test statistics $T_i, i = 1, \ldots, m$, and $Y_i$ is independent of $T_i$ under $H_i$, then
the fixed sequence multiple testing procedures introduced in Theorems 3.1-4.2 can still strongly control the FDR at level $\alpha$.
\end{theorem}

Proof. Assume without any loss of generality that $Y_1 \ge \cdots \ge Y_m$,
so that conditional on the $Y_i$'s, $H_i$ is the $i$th hypotheses to
be tested in our fixed sequence multiple testing methods. When $H_i$
is true, $P_i$ is independent of both $Y_i$ and
$X_j, j = 1, \ldots, m$ with $j \neq i$. This follows from independence of the $X_i$'s and that of $Y_i$ and $T_i$ under $H_i$. Thus, conditional on the
$Y_i$'s, each true null $p$-value $P_i$ still satisfies (1) and
is independent of all other $p$-values $P_j$ with $j \neq i$. Therefore, we have for each of the
procedures in Theorems 3.1, 3.2, 4.1, and 4.2,
\begin{equation}
\expv{\frac{V}{\max(R,1)} \cond Y_1, \ldots, Y_m} \le \alpha.
\label{EQN_DATA_DRIVEN}
\end{equation}
This proves the desired result. \qed

\vspace{10pt}
We applied our proposed methods to the HIV microarray data
(van’t Wout et al., 2003) used by Efron (2008). These data consist
of $m = 7680$ gene expression levels across eight subjects, four HIV
infected and four uninfected. The data were log-transformed and
normalized. Our goal is to determine which genes are differentially
expressed by testing $H_i: \mu_{i}^{(1)} = \mu_{i}^{(2)}$ versus
$H_i': \mu_{i}^{(1)} \neq \mu_{i}^{(2)}$ simultaneously for $i = 1,
\dots, 7680$, where $\mu_{i}^{(1)}$ and $\mu_{i}^{(2)}$ are the gene
specific mean expressions for HIV infected and uninfected subjects,
respectively.

We applied our proposed procedures with the $p$-values generated
from two sample $t$-tests for the genes. Since there is no natural
ordering among the genes, we used the ordering statistics for two
sample $t$-tests in Example 2 to order these tested hypotheses. We
compared the procedure in Theorem
\ref{THM_INDEPENDENCE_FIX_SEQ_WITH_ACC} with
the BH procedure. The results are summarized in Table \ref{TABLE_MICROARRY_2} for
different values of $k$ where $k/m = 0.05, 0.1$, and $0.15$.  As seen from Table \ref{TABLE_MICROARRY_2}, for all values of $k$ except $k=1$, the procedure in Theorem
\ref{THM_INDEPENDENCE_FIX_SEQ_WITH_ACC} generally has more
rejections than the BH procedure. When $\alpha$ is small, $k/m = 0.05$ tends to have the most rejections, but for large $\alpha$, $k/m = 0.1$ has the most rejections. Also, we compared
the procedure in Theorem \ref{THM_ANY_JOINT_FIX_SEQ_WITH_ACC} with the BY procedure. The results are
displayed in Table \ref{TABLE_MICROARRY}.  As seen from Table \ref{TABLE_MICROARRY}, for
most values of $k$, our procedure outperforms the BY procedure in
terms of the number of rejections. When $\alpha = 0.001$, the BY procedure cannot reject any hypotheses, but the procedure in Theorem \ref{THM_ANY_JOINT_FIX_SEQ_WITH_ACC} has at least 8 rejections for all the values of $k$ considered.

\begin{table}[t]
\centering
  \caption{The Number of Discoveries out of $m = 7680$ Genes in the HIV Data from van’t Wout et al. (2003) by the procedure from Theorem \ref{THM_INDEPENDENCE_FIX_SEQ_WITH_ACC} and the BH Procedure}
    \vspace{5pt}
\begin{tabular}{|c|c|c|c|c|c|}
    \hline
    ~             & \multicolumn{4}{c|}{Procedure from Theorem \ref{THM_INDEPENDENCE_FIX_SEQ_WITH_ACC}} & BH Procedure \\ \hline
    ~             & k = 1 & k/m = 0.05 & k/m = 0.1 & k/m = 0.15 & ~ \\ \hline \hline
    $\alpha = 0.001$ &11   &13   &9   &8   &8    \\ \hline
    $\alpha = 0.01$ &11   &18   &17   &16   &13    \\ \hline
    $\alpha = 0.025$ &11   &18   &18   &18   &13    \\ \hline
    $\alpha = 0.05$ &11   &20   &19   &19   &18    \\ \hline
    $\alpha = 0.1$ &20   &21   &24   &20   &22    \\ \hline
\end{tabular}
    \label{TABLE_MICROARRY_2}
\vspace{15pt}
    \caption{The Number of Discoveries out of $m = 7680$ Genes in the HIV Data from van’t Wout et al. (2003) by the procedure from Theorem \ref{THM_ANY_JOINT_FIX_SEQ_WITH_ACC} and the BY Procedure}
    \vspace{5pt}
\begin{tabular}{|c|c|c|c|c|c|}
    \hline
    ~             & \multicolumn{4}{c|}{Procedure from Theorem \ref{THM_ANY_JOINT_FIX_SEQ_WITH_ACC}} & BY Procedure \\ \hline
    ~             & k = 1 & k/m = 0.05 & k/m = 0.1 & k/m = 0.15 & ~ \\ \hline \hline
    $\alpha = 0.001$ &11   &10   &8   &8   &0    \\ \hline
    $\alpha = 0.01$ &11   &13   &13   &11   &8    \\ \hline
    $\alpha = 0.025$ &11   &15   &13   &13   &10    \\ \hline
    $\alpha = 0.05$ &11   &16   &15   &13   &10    \\ \hline
    $\alpha = 0.1$ &11   &18   &16   &16   &13    \\ \hline
\end{tabular}

    \label{TABLE_MICROARRY}
\end{table}

\section{Simulation Study}

A simulation study was conducted to address the performances of the
proposed procedures. We will refer to the procedures in Theorems
\ref{THM_ANY_JOINT_FIX_SEQ_NO_ACC},
\ref{THM_INDEPENDENCE_FIX_SEQ_NO_ACC},
\ref{THM_ANY_JOINT_FIX_SEQ_WITH_ACC}, and
\ref{THM_INDEPENDENCE_FIX_SEQ_WITH_ACC} as Procedures 1-4,
respectively. Specifically, we addressed the following two
questions:
\begin{enumerate}
    \item How do Procedures 1, 2, 3, and 4 compare against the BH and BY procedures in terms of FDR and power?
    \item For Procedures 3 and 4, how should $k$ be chosen so that the power is large?
\end{enumerate}

\begin{figure}[t!]
    \begin{center}
        \includegraphics[width=1\textwidth, angle=-90]{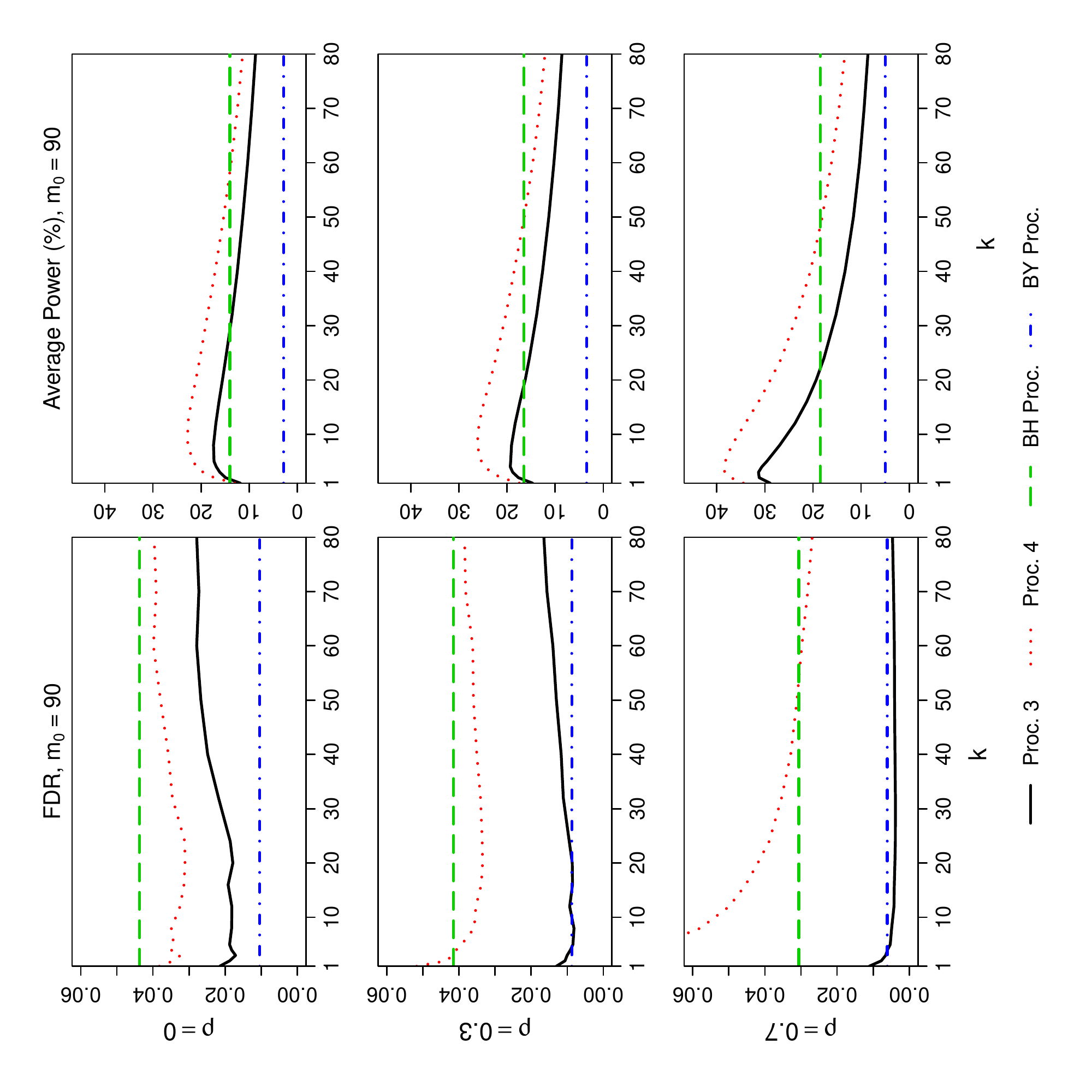}
        \vskip -20pt
        \caption{The FDR (left) and average power (right) of Procedure 3 (solid), Procedure 4 (dotted), the BH procedure (dashed), and the BY procedure (dotted dash) for 100 hypotheses with $m_0 = 90$ and $n = 10$ under common correlation.}
       \label{IMG_PLOT_90}
    \end{center}
\end{figure}

In each simulation, $n$ independent $m$ dimensional random normal
vectors with covariance matrix $\Sigma$ and components $Z_i \sim
N(\mu_i, 1), i = 1, \dots, m$, were generated. The $p$-value for
testing $H_i: \mu_i = 0$ vs. $H'_i: \mu_i > 0$ was calculated using
a one-sided, one-sample $t$-test for each $i$. The $\mu_i$ corresponding to
each false null hypothesis is set to the value at which the power of
one-sample t-test at level $0.05$ is $0.75$.  As for the joint dependence, we considered a common correlation structure where $\Sigma$ had off-diagonal components equal to $\rho$ and diagonal components equal to 1.

We set $\alpha = 0.05$ and $m = 100$. The hypotheses were ordered using the `sum of squares ordering' used in Example 1 from Section 5. We had 5,000 runs of simulation for each of the procedures considered. We noted the false discovery proportion and the proportion of correctly rejected false null hypotheses for each procedure in each of these runs. The simulated FDR and average power (the expected proportion of correctly rejected false null hypotheses) were obtained by averaging out the corresponding 5,000 values.

\begin{figure}[t!]
    \begin{center}
        \includegraphics[width=1\textwidth, angle=-90]{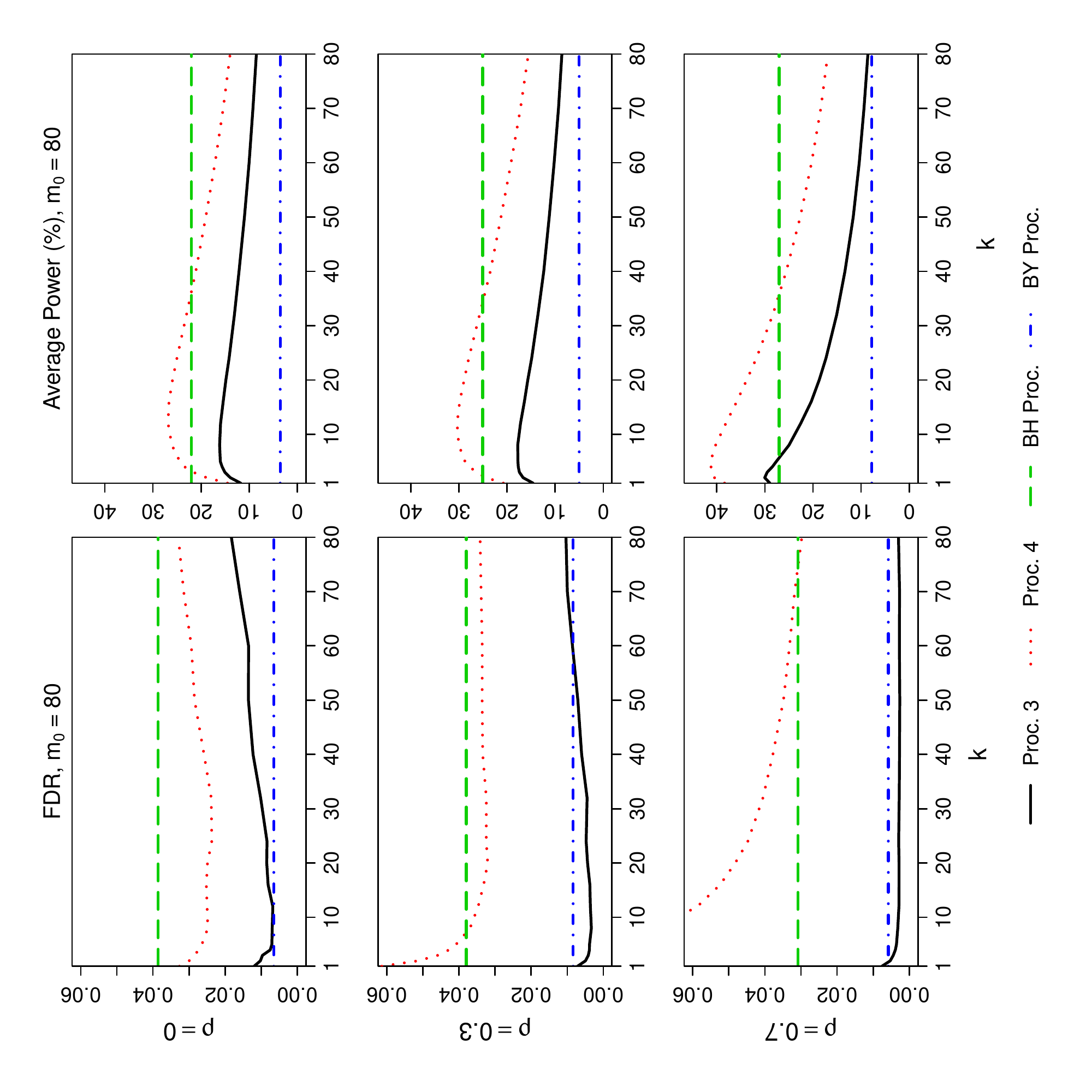}
        \vskip -20pt
        \caption{The FDR (left) and average power (right) of Procedure 3 (solid), Procedure 4 (dotted), the BH procedure (dashed), and the BY procedure (dotted dash) for 100 hypotheses with $m_0 = 80$ and $n = 10$ under common correlation.}
       \label{IMG_PLOT_80}
    \end{center}
\end{figure}

We first compared Procedures 1-4 with the BH and BY procedures when $m_0 = 90$ and $n = 10$. Figure \ref{IMG_PLOT_90} displays these comparisons in terms of the simulated FDR and average power, respectively, under common correlation with $\rho = 0$ (independence), $\rho = 0.3$, and $\rho = 0.7$ while $k$ varies from 1 to 80. We did not explicitly label Procedures 1 and 2 as these are both special cases of Procedures 3 and 4 when $k = 1$, respectively. As seen from Figure \ref{IMG_PLOT_90}, Procedure 3 controls the FDR at level $0.05$ under every dependence configuration. Procedure 4 controls the FDR under independence ($\rho = 0$), generally maintains controls of the FDR under mild correlation ($\rho = 0.3$), but loses control of the FDR for small values of $k$ under strong correlation ($\rho = 0.7$). From Figure \ref{IMG_PLOT_90}, one can also see that Procedure 3 tends to have its largest power when $k$ is between about 3 and 10, and Procedure 4 tends to have its largest power when $k$ is between about 5 and 15. In addition, Figure \ref{IMG_PLOT_90} shows that for a well chosen $k$, Procedures 3 and 4 outperform the BH procedure, and for all values of $k$, Procedures 3 and 4 outperform the BY procedure.

Next, we compared the FDR and average power of these procedures when $m_0 = 80$ and $n = 10$. Figure \ref{IMG_PLOT_80} displays these comparisons for $\rho = 0$, $\rho = 0.3$, and $\rho = 0.7$ while $k$ varies from 1 to 80. In terms of FDR, the results shown in Figure \ref{IMG_PLOT_80} are similar to the results shown in Figure \ref{IMG_PLOT_90}. In terms of power, Figure \ref{IMG_PLOT_80} shows that the powers of all the procedures tend to show an increase over the powers shown in Figure \ref{IMG_PLOT_90}. Figure \ref{IMG_PLOT_80} also shows that for a well chosen $k$, Procedures 4 can outperform the BH procedure, but, in general, Procedure 3 cannot. Again, for all values of $k$, Procedures 3 and 4 outperform the BY procedure.

\section{Concluding Remarks}
In this paper, we have developed `step-down' procedures which control the FDR and
exploit the structure of pre-ordered hypotheses. We have been able
to produce the desired methods in the most simple as well as a
general setting covering different dependence scenarios. Our
simulation study and real data analysis show that in some cases,
the proposed procedures can be powerful alternatives to
existing FDR controlling procedures.

Using some of the techniques developed in this paper, it is possible
to develop other types of fixed sequence procedures controlling the
FDR, such as a fallback-type procedure. Unlike the conventional and
generalized fixed sequence procedures developed in this paper, the
fallback-type procedure tests the remaining hypotheses no matter how
many earlier hypotheses are accepted, which is needed for analyzing
stream data in sequential change detection problems.

Although we have only considered the simplest hierarchical structure
-- fixed sequence structure -- by using the similar techniques
presented in this paper,  we were able to develop simple and
powerful procedures that control the FDR under various dependencies
when testing multiple hypotheses with a more complex hierarchical
structure. We plan to present these procedures in a future
communication.



\vskip 10pt

\section*{Appendix}
\subsection{Proof of Theorem \ref{THM_ANY_JOINT_FIX_SEQ_NO_ACC} (ii)}

For any $u_1$, $1 \le u_1 \le m$, consider a joint distribution of
the $p$-values such that the first $u_1-1$ hypotheses are false null
hypotheses whose corresponding $p$-values are $0$ with probability
one. The remaining $m-u_1+1$ hypotheses are true null hypotheses
such that $\hat{P}_1 \sim U(0,1)$ and $\hat{P}_i = \hat{P}_1$ for $i
= 2, \dots, m_0$.  Under such joint distribution of the $p$-values,
the FDR of the conventional fixed sequence procedure is exactly
$\critA{u_1}(m-u_1+1)/m$. If $\critA{u_1} = 1$ then the critical
constant is already at its largest and cannot be improved.
Otherwise, if $\critA{u_1} < 1$, then FDR $=\alpha$ and thus
$\critA{u_1}$ cannot be made any larger.

In the above construction $\vec{P}$ is PRDS on $\vec{P}_0$. We only
need to note that for $i = 1, \ldots, m_0$, $\widehat P_i = p$
implies $\vec{P} = (0, \dots, 0, p, \dots, p)$. Thus, for any
increasing set $D$,
\begin{align*}
\pr{\vec{P} \in D | \widehat P_i = p} =
\begin{cases}
1 &\text{ if } (0, \dots, 0, p, \dots, p ) \in D \\
0 &\text{ otherwise},
\end{cases}
\end{align*}
which is an increasing function in $p$. \qed

\subsection{Proof of Lemma \ref{LEMMA_FDR_PROCESS_SPECIAL}}
Lemma \ref{LEMMA_FDR_PROCESS_SPECIAL} can be regarded as a special
case of Lemma \ref{LEMMA_FDR_PROCESS} with $k = 1$. Note that
for $i = 1, \ldots, m$, $m_{1,i} = m_{1,i-1}$ and
$m_{0,i} = m_{0,i-1} + 1$ when $i \in I_0$,
$m_{0,i} = m_{0,i-1}$ when $i \notin I_0$, and
$m_{0,i-1} + m_{1,i-1} = i-1$. Thus, when $k = 1$, the event $\{R \ge i \}$ implies
$V_i = m_{0,i}$ and hence
\begin{equation}
\frac{V_i}{i} - \frac{V_{i-1}}{i-1} =
\begin{cases}
\ind{1 \in I_0} &\text{ for } i = 1\\
\frac{m_{1,i-1}\ind{i \in I_0} - m_{0,i-1}\ind{i \notin
I_0}}{i(i-1)} &\text{ for } i = 2, \dots, m. \label{EQN_LEMMA_DI}
\end{cases}
\end{equation}
By (\ref{EQN_LEMMA_DI}) and Lemma \ref{LEMMA_FDR_PROCESS}, the desired result follows. \qed

\subsection{Proof of Lemma \ref{LEMMA_FDR_PROCESS}}
It is easy to see that
\begin{eqnarray*}
     \text{FDR} & = & \expv{\csum{i = 1}{m}\frac{V_i}{i}\ind{R = i}}
= \expv{\csum{i = 1}{m}\left(\frac{V_i}{i}\ind{R \ge i} - \frac{V_i}{i}\ind{R \ge i+1}\right)} \\
& = & \expv{\csum{i =
1}{m}\left(\frac{V_i}{i}-\frac{V_{i-1}}{i-1}\right)\ind{R \ge i}}\\
& = & \expv{\csum{i =
1}{m}\left(\frac{V_i}{i}-\frac{V_{i-1}}{i-1}\right)\ind{J_i < m+1}},
\end{eqnarray*}
the desired result. ~\qed

\subsection{Proof of Lemma \ref{LEMMA_INDEP2}}
It is easy to see that for $i=1, \ldots, m$, if there are at least $i$ rejections , then $i \le J_i \le \min(i+k-1, m)$. For ease of
notation, let $\maxj{i} = \min(i+k-1, m)$. For $i, j = 1, \ldots, m$,
define $f_i(j) = \frac{(k-j+i)\alpha}{k}\frac{S_i}{i}$ and  $W_i(j)
= \ind{J_{i-1} \le j, J_i > j}$. Regarding the relationship between
$J_i$ and $W_i(j)$, there are the following two equalities
available:
\begin{align}
\ind{J_i = j} =  W_i(j-1)\ind{P_j \le \critD{j}}
\label{LEMMA_INDEP2_EQN2}
\end{align}
and
\begin{eqnarray}
& & W_i(j)-W_i(j-1) =  \ind{J_{i-1} = j} - \ind{J_{i} = j}.
\label{LEMMA_INDEP2_EQN3}
\end{eqnarray}
The first equality follows from the fact that for $i = 1, \ldots, m$ and $j = i, \dots, \maxj{i}$, when $J_i = j$, there are $i-1$ rejections among the
first $j-1$ tested hypotheses and the $i^{th}$ rejection is exactly
the $j^{th}$ tested hypothesis, thus
\begin{align}
\ind{J_i = j} = \ind{J_{i-1} \le j - 1, J_i > j-1, P_j \le
\critD{j}} = W_i(j-1)\ind{P_j \le \critD{j}}. \nonumber
\end{align}
The second equality follows from the fact that the event $\{W_i(j) =
1\}$ implies that there are exactly $i-1$ rejections among the first
$j$ tested hypotheses, thus for $j = i-1, \dots, \maxj{i-1}$,
\begin{eqnarray}
& & W_i(j)-W_i(j-1) = \ind{J_{i-1} \le j, J_i > j}-W_i(j-1) \nonumber \\
& = & \ind{J_{i-1} = j}+\ind{J_{i-1} \le j-1, J_i > j-1, P_j > \critD{j}}-W_i(j-1) \nonumber \\
& = & \ind{J_{i-1} = j} - W_i(j-1)\ind{P_j \le \critD{j}} \nonumber \\
& = & \ind{J_{i-1} = j} - \ind{J_{i} = j}, \nonumber
\end{eqnarray}
where, the third equality follows from the fact that $\ind{P_j >
\critD{j}} = 1 - \ind{P_j \le \critD{j}}$ and the fourth follows
from (\ref{LEMMA_INDEP2_EQN2}).

By using the above two equalities, we can prove two inequalities
below, which are needed in the proof of this lemma. Firstly, we show
by using (\ref{LEMMA_INDEP2_EQN2}) that the following inequality holds:
\begin{align}
\expv{\left(\frac{V_i}{i} - \frac{V_{i-1}}{i-1} + f_i(j) -
f_{i-1}(j)\right)\ind{J_i = j}} \le
\expv{\frac{\alpha}{k}\frac{S_{i-1}}{i-1}W_i(j-1)}.
\label{LEMMA_INDEP2_EQN1}
\end{align}

Proof of (\ref{LEMMA_INDEP2_EQN1}). To see this, we consider, separately, the case
when $j \in I_0$ and when $j \notin I_0$.

Suppose $j \in I_0$, then $S_i = S_{i-1}$ and $V_i = V_{i-1} + 1$ when $J_i = j$.  Using the fact
that $V_{i-1} + S_{i-1} = i-1$, the left hand side of
(\ref{LEMMA_INDEP2_EQN1}), after some algebra, becomes
\begin{eqnarray*}
    & & \expv{\frac{k+(j-k)\alpha}{ki}\frac{S_{i-1}}{i-1}W_i(j-1)\ind{P_j \le \critD{j}}} \\
& = & \expv{\frac{k+(j-k)\alpha}{ki}\frac{S_{i-1}}{i-1}W_i(j-1)}\pr{P_j \le \critD{j}} \\
& \le & \expv{\frac{k+(j-k)\alpha}{ki}\frac{S_{i-1}}{i-1}W_i(j-1)}\frac{i\alpha}{k+(j-k)\alpha} \\
& = & \expv{\frac{\alpha}{k}\frac{S_{i-1}}{i-1}W_i(j-1)}.
\end{eqnarray*}
The first equality follows from these two facts: (i) When $J_i = j$,
i.e. the $i^{th}$ rejection is $H_j$,
$S_{i-1}$ is only dependent on the first $j-1$ $p$-values, since
$S_{i-1}$ is the number of false null hypotheses among
the first $i-1$ rejections; (ii) $W_i(j-1)$ is also only dependent on
the first $j-1$ $p$-values, since $W_i(j-1)$ is 1 if and only if
there are exactly $i-1$ rejections among the first $j-1$ hypotheses tested.
The inequality follows from (1).

Now suppose $j \notin I_0$, then
$S_i = S_{i-1}+1$ and $V_i = V_{i-1}$.  Similarly, using the fact that $V_{i-1} + S_{i-1} =
i-1$,  the left hand side of
(\ref{LEMMA_INDEP2_EQN1}), after some algebra, becomes
\begin{eqnarray*}
    & & \expv{\left(\frac{-V_{i-1}}{i(i-1)}+\frac{((j-k)S_{i-1} + (i-1)(k-j+i))\alpha}{ki(i-1)}\right)W_i(j-1)\ind{P_j \le \critD{j}}} \\
 & \le & \expv{\left(\frac{-V_{i-1}}{i(i-1)}\frac{(k-j+i)\alpha}{k}+\frac{((j-k)S_{i-1} + (i-1)(k-j+i))\alpha}{ki(i-1)}\right)W_i(j-1)} \\
& = & \expv{\frac{\alpha}{k}\frac{S_{i-1}}{i-1}W_i(j-1)}.
\end{eqnarray*}
The inequality follows by the fact that $j \ge i$ so that $k-j+i \le
k$.  In addition, in the last line we use the fact that $V_{i-1} +
S_{i-1} = i-1$. ~\qed

Next, we show by using (\ref{LEMMA_INDEP2_EQN3}) that the following inequality holds:
\begin{align}
\expv{f_{i-1}(J_{i-1})\ind{J_{i-1} < m+1} - f_{i-1}(J_{i})\ind{J_{i}
< m+1}} \ge
\expv{\csum{j=i}{\maxj{i}}\frac{\alpha}{k}\frac{S_{i-1}}{i-1}W_i(j-1)}.
\label{LEMMA_INDEP2_EQN4}
\end{align}

Proof of (\ref{LEMMA_INDEP2_EQN4}). By using (\ref{LEMMA_INDEP2_EQN3}), we have
\begin{eqnarray*}
& & \qquad \expv{f_{i-1}(J_{i-1})\ind{J_{i-1} < m+1} - f_{i-1}(J_{i})\ind{J_{i} < m+1}} \\
& = & \expv{\csum{j=i-1}{\maxj{i-1}}f_{i-1}(j)\ind{J_{i-1} = j} - \csum{j=i}{\maxj{i}}f_{i-1}(j)\ind{J_i = j}} \\
& = & \expv{\csum{j=i-1}{\maxj{i-1}}f_{i-1}(j)\ind{J_{i-1} = j} - \csum{j=i}{\maxj{i-1}}f_{i-1}(j)\ind{J_i = j}} \\
& = & \expv{\csum{j=i}{\maxj{i-1}}f_{i-1}(j)\left(\ind{J_{i-1} = j} - \ind{J_{i} = j}\right) + f_{i-1}(i-1)\ind{J_{i-1} = i-1}} \\
& = & \expv{\csum{j=i}{\maxj{i-1}}f_{i-1}(j)(W_i(j) - W_i(j-1)) + f_{i-1}(i-1)W_i(i-1)} \\
& = & \expv{\csum{j=i}{\maxj{i-1}}\left(f_{i-1}(j-1) - f_{i-1}(j)\right)W_i(j-1) + f_{i-1}(\maxj{i-1})W_i(\maxj{i-1})} \\
& \ge &
\expv{\csum{j=i}{\maxj{i}}\frac{\alpha}{k}\frac{S_{i-1}}{i-1}W_i(j-1)},
\end{eqnarray*}
the desired result. Here, the second equality follows from the fact that if $\maxj{i-1} = m$,
then $\maxj{i} = m$; otherwise, $\maxj{i-1} = i+k-2$ and $\maxj{i} =
i+k-1$ so that $f_{i-1}(\maxj{i}) = 0$.  The fourth equality follows
from (\ref{LEMMA_INDEP2_EQN3}) and the fact that $W_i(i-1) =
\ind{J_{i-1} = i-1}$. The inequality follows from the definition of
$f_{i-1}(j)$. ~\qed

Proof of Lemma \ref{LEMMA_INDEP2}. Finally, by combining these two
inequalities, we can get the desired result as follows.
\begin{eqnarray*}
    & & \qquad \expv{\left(\frac{V_i}{i}-\frac{V_{i-1}}{i-1}\right)\ind{J_i < m+1}} \\
& = & \expv{\left(\frac{V_i}{i}-\frac{V_{i-1}}{i-1} + f_i(J_i) - f_{i-1}(J_i)\right)\ind{J_i < m+1}} \\
    & &- \expv{f_{i-1}(J_{i-1})\ind{J_{i-1} < m+1} - f_{i-1}(J_{i})\ind{J_{i} < m+1}} \\
    & &+ \expv{f_{i-1}(J_{i-1})\ind{J_{i-1} < m+1} - f_i(J_i)\ind{J_i < m+1}} \\
& = & \expv{\csum{j=i}{\maxj{i}}\left(\frac{V_i}{i}-\frac{V_{i-1}}{i-1} + f_i(j) - f_{i-1}(j)\right)\ind{J_i = j}} \\
    & &- \expv{f_{i-1}(J_{i-1})\ind{J_{i-1} < m+1} - f_{i-1}(J_{i})\ind{J_{i} < m+1}} \\
    & &+ \expv{f_{i-1}(J_{i-1})\ind{J_{i-1} < m+1} - f_i(J_i)\ind{J_i < m+1}} \\
& \le &\expv{\csum{j=i}{\maxj{i}}\frac{\alpha}{k}\frac{S_{i-1}}{i-1}W_i(j-1) - \csum{j=i}{\maxj{i}}\frac{\alpha}{k}\frac{S_{i-1}}{i-1}W_i(j-1)} \\
    & &+ \expv{f_{i-1}(J_{i-1})\ind{J_{i-1} < m+1} - f_i(J_i)\ind{J_i < m+1}} \\
& = &\expv{f_{i-1}(J_{i-1})\ind{J_{i-1} < m+1} - f_i(J_i)\ind{J_i <
m_1}},
\end{eqnarray*}
The inequality follows from (\ref{LEMMA_INDEP2_EQN1}) and (\ref{LEMMA_INDEP2_EQN4}). ~\qed

\end{document}